\documentclass{PoS}

\usepackage{graphicx,graphics}

\usepackage{dcolumn}
\usepackage{amsmath}
\usepackage{array}
\usepackage{bm}
\usepackage{amssymb}
\usepackage{amsfonts}
\usepackage{color}

\title{Relevance of glueball bound states in the Yang-Mills plasma within a
many-body $T$-matrix approach}

\ShortTitle{Glueball states in the YM plasma within $T$-matrix approach}

\author{\speaker{Daniel Cabrera}\thanks{The author acknowledges kind support
from Service de Physique Nucl\'{e}aire et Subnucl\'eaire,
Universit\'{e} de Mons, where this work was partly done.}\\
        Departamento de F\'{\i}sica Te\'orica II, Universidad Complutense, 28040
	Madrid\\
        E-mail: \email{daniel.cabrera@fis.ucm.es}}

\author{Gwendolyn Lacroix\\
        Service de Physique Nucl\'{e}aire et Subnucl\'eaire,
Universit\'{e} de Mons -- UMONS, Place du Parc 20, 7000 Mons, Belgium\\
        E-mail: \email{gwendolyn.lacroix@umons.ac.be}}

\author{Claude Semay\\
        Service de Physique Nucl\'{e}aire et Subnucl\'eaire,
Universit\'{e} de Mons -- UMONS, Place du Parc 20, 7000 Mons, Belgium\\
        E-mail: \email{claude.semay@umons.ac.be}}

\author{Fabien Buisseret\\
        Service de Physique Nucl\'{e}aire et Subnucl\'{e}aire,
Universit\'{e} de Mons -- UMONS,
Place du Parc 20, 7000 Mons, Belgium;\\ 
Haute Ecole Louvain en Hainaut (HELHa), Chauss\'ee de Binche 159, 7000 Mons,
Belgium\\
        E-mail: \email{fabien.buisseret@umons.ac.be}}

\abstract{
The strongly coupled phase of Yang-Mills plasma with gauge group $SU(3)$ is
studied in a $T$-matrix approach. The existence of lowest-lying glueballs,
interpreted as bound states of two transverse gluons
(quasi-particles in a many-body setup), is analyzed in a non-perturbative
scattering formalism with the input of lattice-QCD  static potentials. The
relevance of the singlet and the (colored) octet and $\bf{27}$ channels
at finite temperature is
discussed. We compute the equation of state of the system in Dashen, Ma and
Bernstein's formulation of statistical mechanics and compare to quenched $SU(3)$
lattice data. Further analysis for the
general case of $SU(N)$ is envisaged.
}

\FullConference{Sixth International Conference on Quarks and Nuclear Physics,\\
		April 16-20, 2012\\
		Ecole Polytechnique, Palaiseau, Paris}

\begin{document}

\section{Introduction}

The aim of the present work is to study the effect of two-body interactions in
the thermodynamic features of the Yang-Mills (or gluon) plasma. We take
recourse of  Dashen, Ma and Bernstein's formulation of statistical
mechanics~\cite{dashen} in
terms of the $S$- (or $T$-) matrix operator of gluon-gluon
scattering. Correlations 
 (leading to glueball bound states in the plasma) are studied
within a non-perturbative approach which allows to investigate the behavior of
the system in a range of temperatures where it is strongly
interacting. We start by describing our setup for gluon-gluon interaction
at finite temperature, and then apply our
formalism to calculate the equation of state of the system. Next we discuss
our result for the glueball spectrum in different color channels with special
emphasis in dissociation temperatures and correlations above threshold. Finally,
we compare our calculation of the pressure with recent lattice data in the
pure gauge sector and discuss further applications of the present approach.

\section{Two (quasi-)gluon scattering in a $T-$matrix approach}
We follow Jacob and Wick's helicity formalism~\cite{jaco} in order to
describe a two-gluon state, the gluons being considered as transverse spin-1
bosons.
As shown in \cite{barnes-heli}, two-gluon states can be organized in
four families of helicity states in a $J^P$ basis, 
separated in helicity singlets $\left|S_\pm; J^P\right\rangle$ and doublets
$\left|D_\pm; J^P\right\rangle$ (symmetrization for identical two-particle
states imposes selection rules on  $J$).
Obviously, one cannot include all the possible $J^P$ channels contributing to
the dynamics of the system. A valid criterion to select the most relevant ones
is to keep low on the averaged orbital angular momentum. This 
has been used to interpret
the mass hierarchy of the glueball spectrum~\cite{barnes-heli} (in a simple
nonrelativistic picture, it provides the strength of the 
centrifugal barrier in scattering theory).
With $\left\langle \vec L^2\right\rangle=2$ one has, in terms of the standard
$\left|^{2S+1}L_J\right\rangle$ basis,
\begin{equation}\label{scaps}
    \left|S_+;0^{+}\right\rangle=\left[\frac{2}{3}\right]^{1/2}\left|^1 S_0\right\rangle+\left[\frac{1}{3}\right]^{1/2}\left|^5 D_0\right\rangle,
    \quad
    \left|S_-;0^{-}\right\rangle=-\left|^3P_0\right\rangle,
\end{equation}    
which in the singlet channel correspond to the $0^{++}$ and $0^{-+}$ glueballs,
respectively. With $\left\langle \vec L^2\right\rangle=4$ one has the symmetric
state $\left|D_+;2^{+}\right\rangle$
plus three antisymmetric states: $\left|S_+;1^{-}\right\rangle$,
$\left|S_-;1^{+}\right\rangle$ and $\left|D_-;2^{-}\right\rangle$.
The symmetric
$2^+$ state corresponds to the singlet $2^{++}$ glueball, which is also among
the lightest ones in vacuum. We shall keep no further states in our study.

The dynamics of gluon-gluon scattering in the plasma is encoded by the in-medium
two-body scattering amplitude or $T$-matrix, which satisfies the (3D-reduced) Bethe-Salpeter
equation (in
partial waves)
\begin{equation}\label{T-matrix}
T_{J^P}(E; q,q') = V_{J^P}(q,q') + \displaystyle\frac{1}{(2\pi)^3}
\displaystyle\int_0^\infty dk \,  k^2\, V_{J^P}(q,k) \,G(E;k)\, T_{J^P}(E;k,q'),
\end{equation}
with $q$ ($q'$) the in-coming (out-going) momentum of a gluon in the center of
mass reference frame and $E$ the total energy of the gluon pair. The two-body
interaction potential in momentum space, $V_{J^P}$, is related to the potential
in coordinate space, $V(r)$, by Fourier transform and partial wave projection.
The resolvent function in Eq.~(\ref{T-matrix}) is given by the two-gluon
propagator,
\begin{equation}
G(E;k) = \displaystyle\frac{m_g^2}{\epsilon(k)} \displaystyle\frac{1}{\epsilon(k)^2 - E^2/4 + 2i \, \epsilon(k)\, \Sigma}
\end{equation}
with the gluon dispersion relation $\epsilon(k)= \displaystyle\sqrt{k^2+m_g^2}$
and $\Sigma$ is related to the imaginary part (i.e., decay width) of the
single-particle gluon selfenergy in the plasma.

\section{Equation of state of the Yang-Mills plasma}

The $T$-matrix approach is particularly suited for a
straightforward implementation of the contribution of two-body interactions in
the partition function of the system.
At vanishing chemical potential, the grand potential $\Omega$
of an interacting particle gas is given by \cite{dashen}
\begin{equation}\label{pot0}
\Omega=\Omega_0+\sum_\nu\left[\Omega_\nu-\frac{1}{2\pi^2\beta^2}\int^\infty_{M_\nu}
\frac{d\epsilon}{4\pi i}\, \epsilon^2\,  K_2(\beta\epsilon)\, \left. {\rm
Tr}_\nu \left({\cal S}S^{-1}\overleftrightarrow{\partial_\epsilon}S
\right)\right|_c\right]. 
\end{equation}  
The first term, $\Omega_0$, is the grand potential of a gas of
free relativistic particles (\textit{i.e.}, when the interactions are turned
off). The second term accounts for 
interactions in the plasma. The sum runs on all the quantum numbers
($\nu$)
necessary to fix a scattering channel. 
Below
threshold, the contribution from bound states leads to the
$\Omega_\nu$ term, behaving as
free additional species in the plasma.
The third term accounts for the scattering contribution above threshold.
The operator ${\cal S}$ enforces symmetrization in those channels involving
identical particles, and the subscript $c$ means that only
connected scattering contributions are taken into account.
The scattering operator is defined as $S=1-2\pi i \delta(\epsilon-H_0) T$,
with $H_0$ is the free Hamiltonian of the system. It is worth mentioning that
the scattering contribution sets in at the level of the second order coefficient
in a virial expansion of the partition function. Also note that Eq.~(\ref{pot0}) can be
rewritten, by means of unitarity of the $S$-matrix, in terms of a weighted
thermal average of scattering phase shifts.
Once the off-shell $T$-matrix is known, the pressure is simply given by 
$p=-\Omega$. Other thermodynamic observables can be derived form $p$. For example, the
trace anomaly $\Delta=e-3p$ and the entropy density $s$ read, respectively,
$\Delta=-\frac{1}{\beta^3} \left[ \partial_\beta\left(\beta^4 p\right)\right]
_{V,\,\beta\mu}$ and $s=-\beta^2\left[ \partial_\beta p\right]_{V,\,\mu}$.

\begin{figure}[ht]
\begin{center}
\includegraphics*[width=0.47\textwidth]{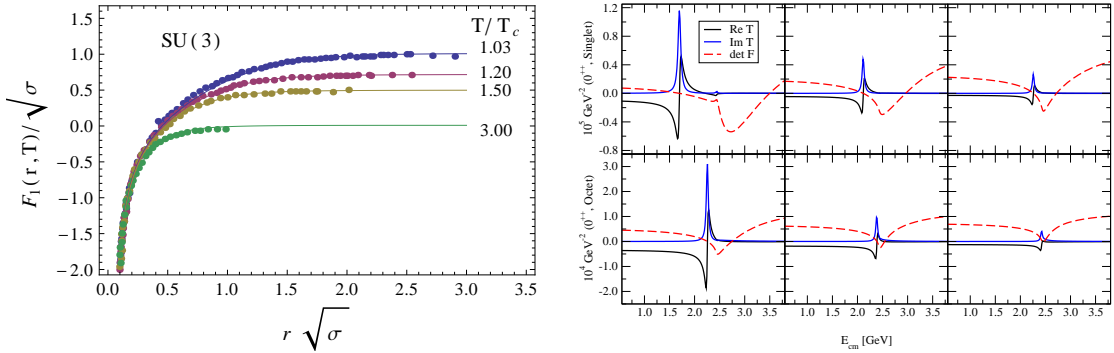}
\hspace{0.3cm}
\includegraphics*[width=0.47\textwidth]{T-0pp-Tevol-S-and-O.eps}
\caption{Left: Static free energy $F_1(r,T)$ of a quark-antiquark pair
bound in a color singlet, computed in SU(3) quenched lattice QCD \cite{kacz3}.
Solid lines are given by an appropriate parameterization \cite{our-paper}.
Right: $T$-matrix for $gg$ scattering in
the $0^{++}$ singlet and octet channels. From left to right
the temperatures are $(1.05, 1.10, 1.15)T_c$.}
\label{fig1}
\end{center}
\end{figure}

\vspace{-.5cm}
\section{Glueballs above $T_c$}
Following the idea in \cite{rapp-cabrera-hess-riek} we take recourse of lattice
QCD simulations to have an estimate of the finite-temperature gluon-gluon
interaction potential from first principles. Accurate computations of the static
free ($F_1$) and internal ($U_1$) energies of a massive quark-antiquark pair in a color singlet
state are available \cite{kacz3}, particularly in quenched $SU(3)$ simulations, which come
handy for our purpose (see Fig.~\ref{fig1}, left panel). Whether it is $F_1$ or $U_1$ (or
none) the most appropriate potential to be identified with the in-medium
two-body interaction at finite temperature
cannot be decided a priori from field-theoretical arguments. Still, as in
\cite{rapp-cabrera-hess-riek}, we choose $U_1$, which typically provides a more
attractive interaction leading to larger dissociation temperatures (and thus
results using this potential should be understood as an upper bound).
We assume color scaling and derive the interaction potential between
two quasi-gluons in color channel ${\cal C}$ as
\begin{equation}\label{Vg}
V(r,T)=\frac{\kappa_{{\cal C};gg}}{\kappa_{q \bar q}} \left[U_1(r,T)-U_1(\infty
, T)\right],
\end{equation}
where $\kappa_{q \bar q}$ is the color factor of the singlet quark-antiquark
pair, and $\kappa_{{\cal C};gg} = (C_2^{\cal C}-2N)/2N$ for $SU(N)$, with
$C_2^{\cal C}$ the quadratic Casimir of the gluon pair in representation $\cal
C$.
As is customary, the potential is normalized to zero as to ensure convergence of
the scattering equation.

The only remaining unknown in our approach is the (effective) quasi-gluon mass.
At finite temperature, it is well accepted that the infinite separation distance
plateau in $U_1$ may be interpreted as an in-medium contribution (selfenergy) to
the gluon mass \cite{mocsy-kaczmarek}. The minimal implementation is a shift
to the bare (zero temperature) gluon mass, which at the same time has to be
fixed, for instance, by reproducing the glueball spectrum in vacuum. 
In terms of the asymptotic value of $U_1$, we propose
the mass shift to be
\begin{equation}\label{mg}
m_g(T)^2=(m_g^0)^2+\delta(T)^2\ , \quad
\delta(T) = \sqrt{\frac{C_g}{C_q}} \frac{U_1(\infty,T)}{2} \ .
\end{equation}
We neglect at the moment the momentum dependence of this correction, which could
be addressed in a full many-body study of the
gluon dispersion relation. $\delta (T)$
scales with the one-body Casimir as suggested by the single-particle
character of the correction (using two-body scaling, as in the potential, leads
to a channel dependent mass correction). The structure of the color factor,
$\sqrt{C_g/C_q}$, leads to the correct large-$N$ limit of the gluon mass
(as it is the case for the Hard-Thermal-Loop result at high temperatures). Also note
that this structure emerges if the gluon selfenergy is calculated from the
$T$-matrix in a Brueckner scheme.
\begin{table}[ht]
\begin{center}
\begin{tabular}{c|rrr}
State & Lattice \cite{glulat} & $T$-matrix & CGQCD \cite{cg0} \\
\hline
$0^{++}$ & 1.73 (5)(8) &  2.17 & 1.98\\
$0^{-+}$ & 2.59 (4)(13) & 2.39  & 2.22 \\
$2^{++}$ & 2.40 (2.5)(12) & 2.34  & 2.42 \\
\end{tabular}
\caption{Masses (in GeV) of the lowest-lying glueball states at zero
temperature. Our results (third column), are compared to the
lattice data of Ref.~\cite{glulat} and to the Coulomb gauge QCD
study in \cite{cg0}.} 
\label{tabglueb}
\end{center}
\end{table}

\vspace{-0.5cm}
First of all, we test that our approach provides reasonable results in vacuum 
and, thus, is a solid starting point for the finite temperature study. We use the
"funnel" potential form with string
breaking setting in at  $V_{sb}\simeq 2$~GeV as suggested by the value of (twice)
the lightest gluelump mass. Our results for the ground states in
$0^{++}$, $0^{-+}$ and $2^{++}$ channels are shown in Table~\ref{tabglueb},
together with lattice results from \cite{glulat} and the Coulomb-gauge QCD
calculation in \cite{cg0} (which formally shares many features with our
$T$-matrix formulation). A
few comments are in order: the parameters of the potential, $\alpha$ and
$\sigma$, are optimized for the best fit of the lattice data for $F_1$ beyond
$T_c$. This justifies the slightly too high value of the energy of the $0^{++}$
glueball. A more precise agreement with the lattice glueball spectrum (including
excited states) can be achieved if tuning the vacuum potential
parameters independently of the finite temperature fit. More details are to be discussed
in \cite{our-paper}.

In Fig.~\ref{fig1} (right panel) we show the $T$-matrix,
at several temperatures, for the singlet and octet scalar glueball channels. In
this plot $m_g$ is set to 1.22~GeV 
(namely, we use $m_g^0\simeq 0.7$~GeV and $\delta=V_{sb}/2$, as in vacuum), thus the
medium effects stem only from the temperature dependence of the potential. The
goal is to illustrate the mechanism of dissociation by color screening: as
temperature is increased, the 
bound states progressively move towards the two-particle threshold and
eventually dissolve by turning to scattering states. The red, dashed curve
represents $\det (1- VG)$ and vanishes exactly at the bound state energy. It
also signals resonant states beyond threshold.
\begin{figure}[ht]
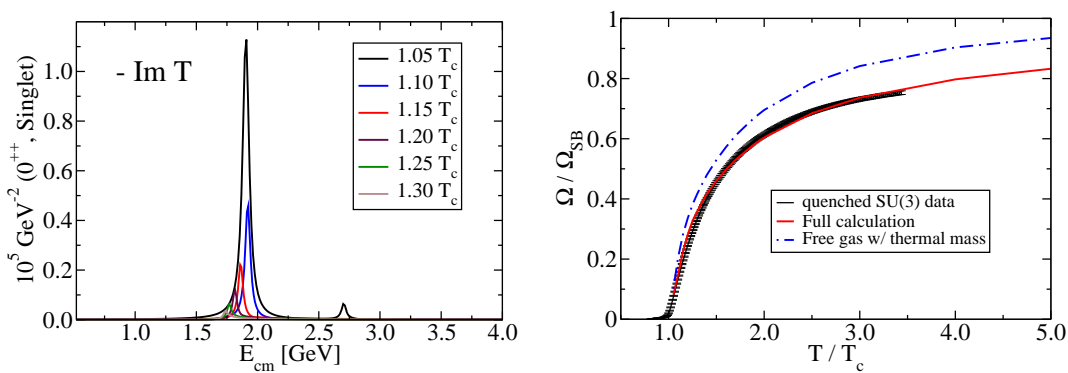

\begin{center}
\includegraphics*[width=0.45\textwidth]{plot-ImT-0ppS-SqrtCasimirM07-Tall.eps}
\hspace{0.3cm}
\includegraphics*[width=0.45\textwidth]{Omega-mg07-SqrtCa-free-plus-full-vs-lattice.eps}
\caption{Left: $\textrm{Im} T$ for $gg$
scattering in the
$0^{++}$ singlet channel with in-medium quasi-gluon masses (see text). Right:
Pressure of the $SU(3)$ gluon-glueball gas. Lattice data from
\cite{Panero}. The dashed curve represents the pressure of the free quasi-gluon
gas (no $gg$ interactions).
}
\label{fig2}
\end{center}
\end{figure}

\vspace{-.5cm}
Next we turn the in-medium gluon mass correction on, $\delta(T)$. The bare gluon
mass parameter is kept to $m_g^0\simeq 0.7$~GeV. This is determined by our best
comparison to the pure gauge lattice QCD data on the pressure, $p=-\Omega$ (see
discussion below).
The results for the imaginary part of the $T$-matrix in the singlet scalar
channel are displayed in Fig.~\ref{fig2} (left panel).
Two competing effects are responsible for the temperature evolution of the
spectrum: reduction of the binding energy and downward shift of the threshold
energy. Overall, the singlet scalar bound state experiences a mild shift to
lower energies and dissociates at about $1.3~T_c$ (this is the value from which
$\det F$ does not vanish anymore; still, considerable strength remains at
threshold up to about $1.5~T_c$). This is in qualitative agreement with the spectral function
analysis of Euclidean correlators by the CLQCD Collaboration \cite{CLQCD}.
We also find bound states in the scalar octet
channel, although considerably less bound. We find  bound states in the
pseudoscalar singlet and octet channels lying right below the threshold energy at
the lowest considered temperature of $1.05~T_c$. The differences between singlet and
octet channels are attributed to the strength of the potential, which scales
with the two-body Casimir. In particular, note that the {\bf 27} channel is
repulsive, and thus no bound states are found. Also note that states in the pseudoscalar
channels, which in our approach correspond to pure $P$-wave scattering, are
just mildly bound due to the centrifugal barrier. The tensor states, having an
$S$-wave component, lie between the scalar and pseudoscalar channels regarding
binding and dissociation temperatures.
%
%
Finally, we compute the partition function of the gluon-glueball gas in the
formalism of Dashen \emph{et al.}, as discussed above. The results, normalized
to the Stefan-Boltzmann pressure, are shown in Fig.~\ref{fig2} (right panel).
Note that, strictly speaking, the only free parameter in the calculation is the
bare gluon mass, $m_g^0$. A value of 0.7~GeV provides a remarkable agreement
with the quenched $SU(3)$ lattice data from Panero~\cite{Panero}. In order to
estimate the effect of two-body interactions, the  partition function for a free
gas including thermal mass corrections is also shown. Note that our bound states
dissolve relatively quickly and only contribute significantly at temperatures at
most up to $\simeq 1.5 T_c$. This illustrates the relevance of accounting for
strong correlations from scattering in the continuum, a contribution which is
readily addressed in our $T$-matrix formalism.

\vspace{-0.2cm}
\section{Summary and outlook}

The relevance of gluon-gluon interactions beyond the critical temperature in the
pure gauge $SU(3)$ plasma has been addressed in a non-perturbative $T$-matrix
many-body framework with the input of (Casimir-scaled) potentials from thermal
lattice QCD. We find glueball bound states in the singlet and octet
channels surviving up to temperatures of about $1.3-1.5 T_c$, together with
sizable threshold effects due to strong correlations beyond the two-particle
threshold. With a minimal number of parameters, we reproduce the equation of
state of the gluon-glueball gas in good agreement with recent quenched $SU(3)$
simulations. 

The $T$-matrix formalism can be applied to calculate bulk thermodynamical
properties of the system such as the sheer viscosity, which can be easily
computed in relaxation-time approximation within a quasi-particle picture.
Our approach also allows for a straightforward extension to $SU(N)$ analysis.
A study of the exceptional group $G_2$, of theoretical interest related to the
origin of (de)\-con\-finement, is also in progress.

\end{document}